\documentclass[aps,reprint,prl,superscriptaddress,longbibliography]{revtex4-1}
\usepackage[version=3]{mhchem}
\usepackage{graphicx}
\usepackage{fancyref}
\usepackage{floatrow}
\usepackage{placeins}
\usepackage{flafter}
\usepackage{soul}
\usepackage{amssymb}
\usepackage{amsmath}
\usepackage{commath}
\usepackage{blindtext}
\usepackage{siunitx}
\usepackage{miller}
\usepackage{xcolor}
\usepackage{float}
\usepackage{lipsum}
\usepackage{bm}
\usepackage{ulem}
\usepackage{physics}
\makeatletter
\let\Hy@backout\@gobble
\makeatother
\usepackage{titlesec}
\titlespacing*{\section}{0pt}{1.1\baselineskip}{\baselineskip}
\setcitestyle{super}

\begin{document}

\author{D.~Davidovikj$^\ddagger$}
\email{d.davidovikj@tudelft.nl}
\affiliation{Kavli Institute of Nanoscience, Delft University of Technology, P.O. Box 5046, 2600 GA Delft, the Netherlands}

\author{D.~J.~Groenendijk$^\ddagger$}
\email{d.j.groenendijk@tudelft.nl}
\affiliation{Kavli Institute of Nanoscience, Delft University of Technology, P.O. Box 5046, 2600 GA Delft, the Netherlands}

\author{A.~M.~R.~V.~L.~Monteiro}
\author{A.~Dijkhoff}
\author{D.~Afanasiev}
\author{H.~S.~J.~van der Zant}
\affiliation{Kavli Institute of Nanoscience, Delft University of Technology, P.O. Box 5046, 2600 GA Delft, the Netherlands}
\author{Y.~Huang}
\author{E.~van~Heumen}
\affiliation{Van der Waals - Zeeman Institute, Institute of Physics (IoP), University of Amsterdam, Science Park 904, 1098 XH Amsterdam, the Netherlands}
\author{A.~D.~Caviglia}
\affiliation{Kavli Institute of Nanoscience, Delft University of Technology, P.O. Box 5046, 2600 GA Delft, the Netherlands}

\author{P.~G.~Steeneken}
\affiliation{Kavli Institute of Nanoscience, Delft University of Technology, P.O. Box 5046, 2600 GA Delft, the Netherlands}
\affiliation{Department of Precision and Microsystems Engineering, Delft University of Technology, Mekelweg 2, 2628 CD Delft, the Netherlands}

\date{\today \\
\vspace{4pt}
\textit{$^\ddagger\text{These authors contributed equally to this work}$}}

\title{Ultrathin complex oxide nanomechanical resonators}

\begin{abstract}
Complex oxide thin films and heterostructures exhibit a profusion of exotic phenomena, often resulting from the intricate interplay between film and substrate. Recently it has become possible to isolate epitaxially grown single-crystalline layers of these materials, enabling the study of their properties in the absence of interface effects. In this work, we create ultrathin membranes of strongly correlated materials and demonstrate top-down fabrication of nanomechanical resonators made out of \ce{SrTiO3} and \ce{SrRuO3}. Using laser interferometry, we successfully actuate and measure the motion of the nanodrum resonators. By measuring their temperature-dependent mechanical response, we observe signatures of structural phase transitions in \ce{SrTiO3}, which affect the strain and mechanical dissipation in the resonators. This approach can be extended to investigate phase transitions in a wide range of materials. Our study demonstrates the feasibility of integrating ultrathin complex oxide membranes for realizing nanoelectromechanical systems on arbitrary substrates.
\end{abstract}
\maketitle
\label{sto}


\section{Introduction}

It is well established that the electronic and magnetic properties of complex oxides are extremely sensitive to mechanical strain due to the strong coupling between the lattice and the charge, spin, and orbital degrees of freedom~\cite{dagotto2005complexity, reyren07,noheda14,holsteen14, zubko16, manca17}. This sensitivity stems from rotations and distortions of the corner-connected \ce{BO6} octahedra (where B is a transition metal ion situated in the centre of the octahedron formed by the oxygen atoms), which determine the overlap between orbitals on adjacent atomic sites~\cite{rondinelli2012control}. The B--O bond lengths and rotation angles are routinely controlled by strain through heteroepitaxy, which forms a powerful tool to tune the properties of ultrathin films. The strong dependence of their electronic properties on mechanical strain has attracted a lot of attention towards their implementation in nanoelectromechanical sensors and actuators~\cite{bhaskar16}, but exploiting this trait to the fullest has been limited by the requirement of a substrate for the epitaxial growth. This constrains the possibilities for their mechanical manipulation and integration with electronics and it could not be circumvented until recently, when single-crystal films of complex oxides were successfully released and transferred~\cite{paskiewicz15,lu16}. This sparked a new wave of interest in studying the intrinsic properties of these materials, this time in their isolated, ultrathin form.

On the other hand, a wide variety of mechanical manipulation techniques have been developed for another class of ultrathin materials, the so-called van der Waals materials~\cite{novoselov16}, where weak interlayer bonding enables exfoliation of single- and few-layer films. Their ease of manipulation has enabled the top-down fabrication of a variety of nanomechanical elements, such as suspended membranes and ribbons. This, combined with their flexibility, low mass and remarkable strength, has made them extremely promising candidates for nanomechanical sensing applications~\cite{atalaya10,bunch12,smith13electromechanical,dolleman15}. Conversely, the well-developed field of nanomechanics has established a solid basis for characterising the thermal and mechanical properties of van der Waals materials~\cite{hone08elastic,dolleman17,davidovikj17nonlinear}. In this work, we utilize the fabrication techniques for van der Waals materials to realize ultrathin nanomechanical resonators made out of epitaxially grown single-crystal complex oxide films. We show that these devices can be used to detect signatures of temperature-induced phase transitions in the material, which are related to its intrinsic properties and the configuration and dynamics of structural domains.

\begin{figure*}[ht!]
	\includegraphics[width=17cm]{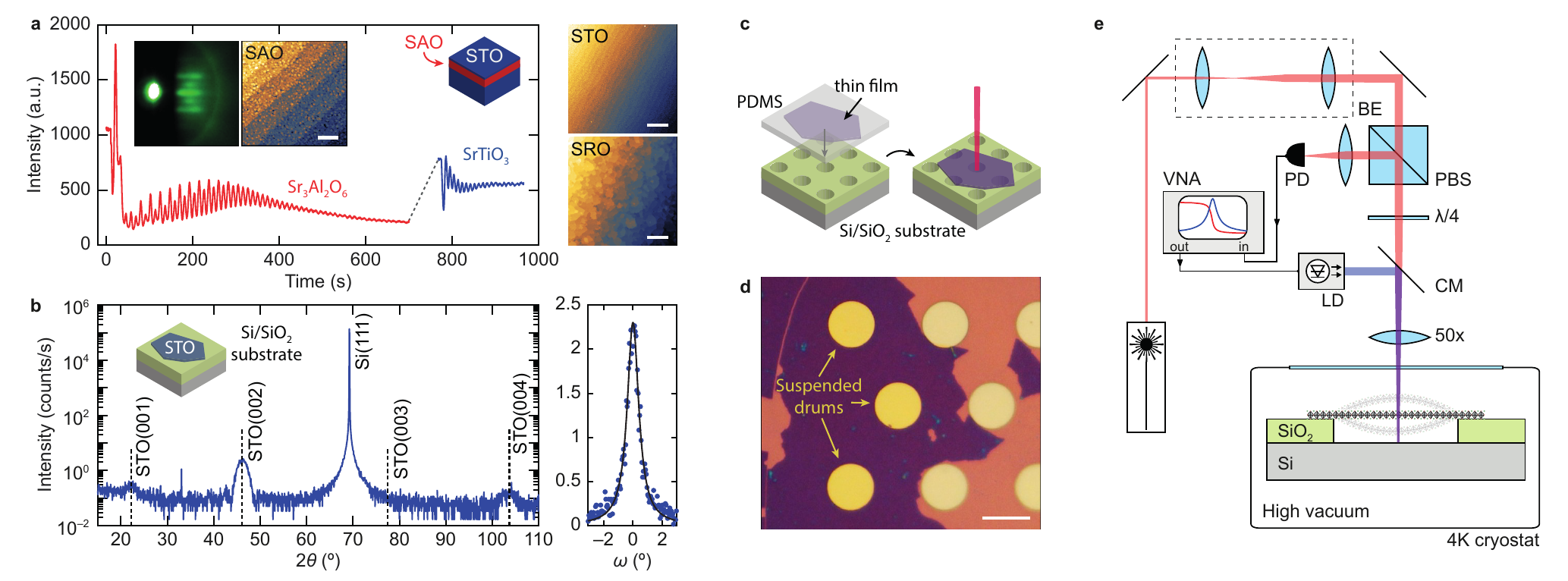}
	\caption{\label{Fig:st-Char} \textbf{Sample fabrication and basic characterisation.} \textbf{a,} RHEED intensity oscillations during the growth of SAO and STO. Inset: RHEED diffraction pattern (left) and AFM image (right) of the SAO surface. The scale bar is 200 nm. Right: AFM images of the STO (top) and SRO (bottom) film surfaces. The scale bar is 1 $\mu$m. \textbf{b,} X-ray diffraction measurement of a 10 unit cell (u.c.) STO film transferred on a \ce{Si}/\ce{SiO2} substrate. Right: rocking curve around the (002) reflection. \textbf{c,} Schematics of the transfer of a thin SRO film onto a pre-patterned \ce{Si}/\ce{SiO2} substrate. \textbf{d,} optical image of suspended (9 u.c.) SRO drums with a diameter of $13\;\mu\mathrm{m}$. The scale bar is 10 $\mu$m. \textbf{e,} Setup for interferometric displacement detection (VNA: Vector Network Analyser, PD: Photodiode, LD: Laser Diode, BE: Beam Expander, PBS: Polarised Beam Splitter, CM: Cold Mirror).} 
\end{figure*}


	
\section{Results}

\subsection{Sample fabrication and characterisation}

The fabrication of the complex oxide mechanical resonators is described in Fig.~\ref{Fig:st-Char}. To isolate the epitaxial \ce{SrTiO3} (STO) and \ce{SrRuO3} (SRO) thin films from the substrate, a water-soluble epitaxial \ce{Sr3Al2O6} (SAO) layer is first deposited by pulsed laser deposition on a \ce{TiO2}-terminated STO(001) substrate (see Methods). Figure~\ref{Fig:st-Char}a shows the reflection high-energy electron diffraction (RHEED) intensity of the specular spot during the growth of SAO and STO. Oscillations are observed during the growth of both films, indicating that the growth occurs in layer-by-layer mode. An atomic force microscopy (AFM) topographic map shows that the STO surface has a step-and-terrace structure, corroborating the growth mode. SRO grows in step-flow mode, as can be inferred from the atomically flat surface and the absence of RHEED oscillations. An XRD measurement of an STO/SAO/SRO heterostructure is shown in Supplementary Figure 1.

To dissolve the sacrificial layer and release the thin film from the substrate, a polydimethylsiloxane (PDMS) layer is attached to the surface before the entire stack is immersed in deionized water. After the dissolution of the SAO layer (approximately 1 hour for a 5 x 5 mm$^2$ 50 nm-thick SAO film, see Supplementary Movie 1), the film can be transferred onto other substrates such as \ce{Si}/\ce{SiO2} using a deterministic dry-transfer technique~\cite{castellanos14}. An X-ray diffraction (XRD) measurement of a 10 u.c.~STO flake on a \ce{Si}/\ce{SiO2} substrate is shown in Fig.~\ref{Fig:st-Char}b (left). Laue oscillations are clearly visible, indicating that the films are of excellent crystalline quality after the release and transfer process. Since the film is no longer epitaxial on the substrate, the rocking curve (right) is a measure of the morphology of the STO film lying on the \ce{SiO2}. The small full width at half maximum ($0.95^\circ$) indicates that the film lies very flat on the \ce{Si}/\ce{SiO2} substrate. Here this growth technique is extended by transferring STO and SRO films onto \ce{Si}/\ce{SiO2} substrates pre-patterned with circular cavities, demonstrating the feasibility of creating suspended complex oxides membranes. An optical image of 9 u.c.~(thickness: $h=$ 3.6 nm) thick SRO drums (diameter: $d= 13~\mu$m) is shown in Fig.~\ref{Fig:st-Char}d. It is remarkable that these materials, much like their van der Waals counterparts, have the flexibility and tensile strength required to be suspended with aspect ratios exceeding $d/h > 3600$.
   
\begin{figure*}[ht!]
		\centering
		\includegraphics[width=12cm]{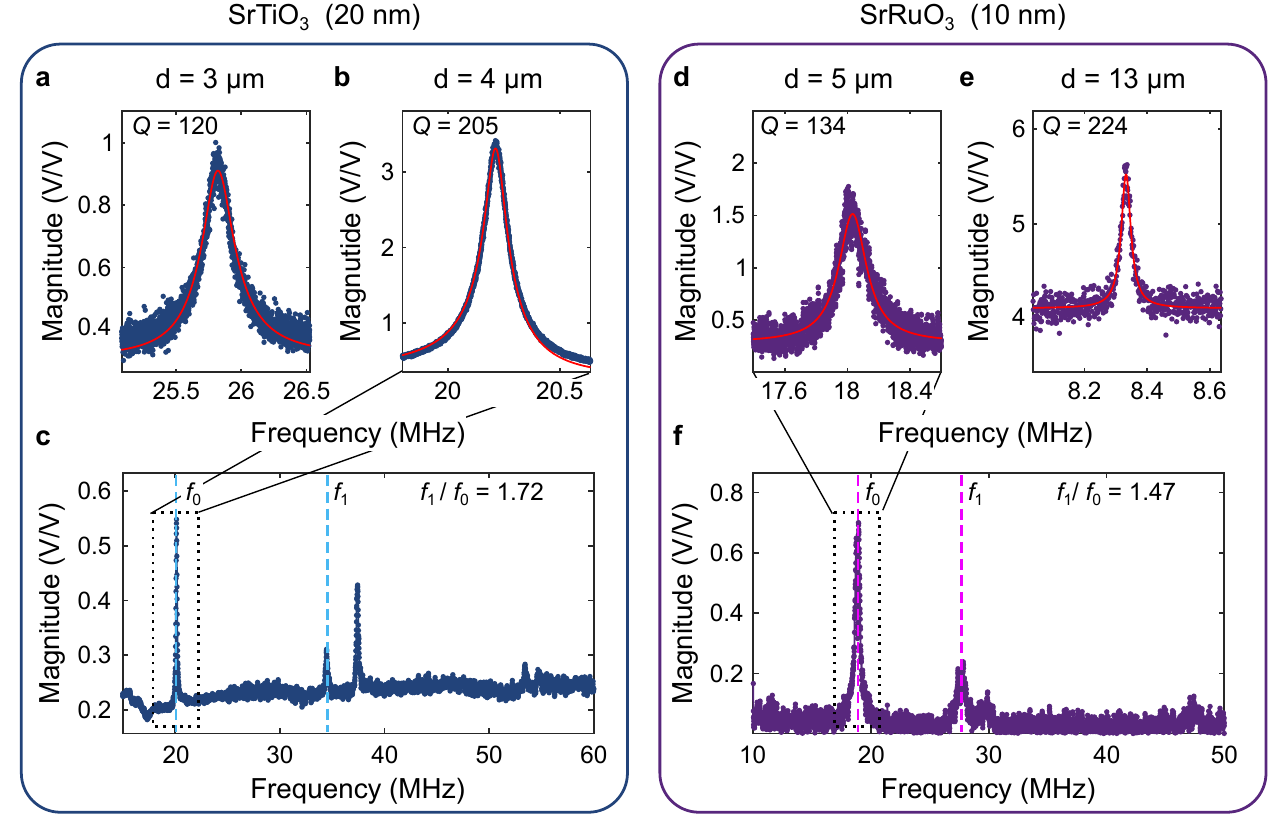}
		\caption{\label{Fig:st-resonancePeaks} \textbf{Mechanical characterisation of STO and SRO nanodrums.} Resonance frequency measurements of 20 nm-thick STO and 10 nm-thick SRO nanodrums. \textbf{a-b,} Frequency spectra of two STO drums with diameters of \textbf{a}: 3 $\mu $m and \textbf{b}: 4 $\mu$m. The red lines are linear harmonic oscillator fits. The extracted quality factors are shown in the top left corners of the panels. \textbf{c,} A wide-range frequency spectrum of the drum shown in \textbf{b}. The positions of the fundamental resonance mode ($f_0$) and the second resonance mode ($f_1$) are marked with vertical dashed lines. \textbf{d-e,} Frequency spectra of two SRO drums with diameters of \textbf{d}: 5 $\mu $m and \textbf{e}: 13 $\mu$m. The red lines are linear harmonic oscillator fits. The extracted quality factors are shown in the top left corners of the panels. \textbf{f,} A wide-range frequency spectrum of the drum shown in \textbf{d}. The positions of the fundamental resonance mode ($f_0$) and the second harmonic ($f_1$) are marked with vertical dashed lines.}
	\end{figure*} 
	
\subsection{Mechanical characterisation of the nanodrums}	
  
We characterise the high-frequency dynamics of the complex oxide nanodrums using the optical actuation and detection setup shown in Fig.~\ref{Fig:st-Char}e. The drums are mounted in the vacuum chamber ($10^{-6}$ mbar) of a closed-cycle cryostat with optical access. Their motion is read out using a red HeNe laser ($\lambda = $ 632.8 nm). The complex oxide membrane and the silicon underneath form a Fabry-P\'{e}rot cavity, where the motion of the membrane modulates the intensity of the reflected light, which is measured by a photodiode. The resonators are actuated optothermally, using a blue laser that is coupled into the optical path via a cold mirror~\cite{bunch07,castellanos13}. Measurements are performed in a homodyne detection scheme using a Vector Network Analyser (VNA), simultaneously sweeping the actuation and detection frequencies.

The mechanical resonances of several STO and SRO drums are shown in Fig.~\ref{Fig:st-resonancePeaks}. Although STO is transparent in the visible range~\cite{cardona65}, the motion of the drums can still be actuated and measured optically since the refractive index of the STO is different from that of vacuum. Figure~\ref{Fig:st-resonancePeaks}a-b shows measurements of two STO drums, and Fig.~\ref{Fig:st-resonancePeaks}d-e of two SRO drums of different diameters. Measurements over a wider frequency range show that higher order resonances of the drums can also be detected; two examples are shown in Fig.~\ref{Fig:st-resonancePeaks}c,f, where up to four higher order resonances are visible. By taking the ratio of the second harmonic to the fundamental mode, we can estimate whether the mechanical properties are dictated by the pre-tension (theoretical ratio 1.59) or if they are dominantly determined by the bending rigidity (theoretical ratio 2.09), the latter being  dependent on the Young's modulus of the material ($E$). It can be seen from Fig.~\ref{Fig:st-resonancePeaks}c that the STO drums are in a cross-over regime (ratio 1.72), similar to what has been observed in drums of similar dimensions made of \ce{MoS2}~\cite{castellanos13} and \ce{TaSe2}~\cite{cartamil15} (an atomic force microscopy nanoindentation measurement of the sample characterised in Fig.~\ref{Fig:st-resonancePeaks}c is shown in Supplementary Figure 2). On the other hand, the mechanical properties of the SRO drums are almost entirely determined by their pre-tension since $f_1/f_0 = 1.47$, which is close to the theoretical value of 1.59. Statistics on 18 STO drums are shown in Supplementary Figure 3. 

\begin{figure*}[t!]
	\centering
	\includegraphics[width=12cm]{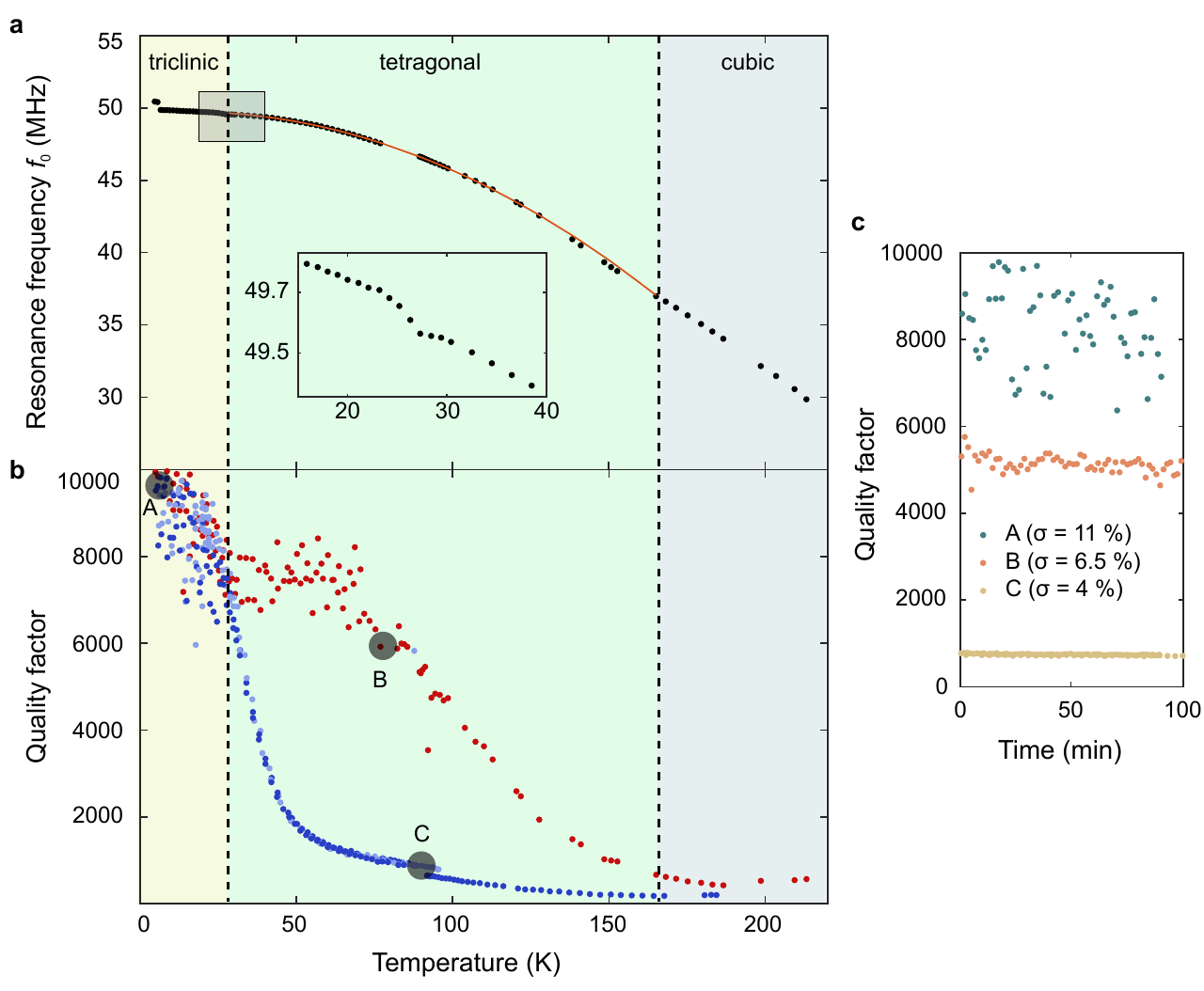}
	\caption{\label{Fig:st-tempSweep} \textbf{Temperature dependence of the mechanical properties of a 4 $\mu$m STO drum.} \textbf{a,} Resonance frequency as a function of temperature (upward sweep). The red line is a guide to the eye in the tetragonal regime. The inset shows a zoomed-in region in the range between 15 and 40 K. \textbf{b,} Quality factor as a function of temperature for seven alternating temperature sweeps. Two distinct branches are distinguishable in the region between 30 and 165 K (dashed lines), coloured in red and blue. The light blue points correspond to cooling cycles. \textbf{c,} Quality factor as a function of time at constant temperatures, where A, B, and C correspond to the points shown in \textbf{b}.}
\end{figure*}

\subsection{Temperature-dependent mechanical properties}

Having confirmed that the resonators can be mechanically characterised at room temperature, we now investigate how their mechanical properties change with temperature. STO is known to undergo a number of phase transitions as a function of temperature, which are expected to influence the mechanical properties of the resonators. Figure~\ref{Fig:st-tempSweep} shows the temperature dependence of the mechanical properties of an STO nanodrum. The temperature dependence of the resonance frequency (Fig.~\ref{Fig:st-tempSweep}a) shows an evolution which is commonly observed in 2D materials~\cite{chen09,singh10,morell16}. This behavior is usually ascribed to a difference in thermal expansion coefficient between the membrane and the substrate~\cite{singh10,morell16}, which results in thermally induced tensile stress. Since the resonance frequency is related to the thermal expansion coefficient ($f_0^2 \propto \alpha_\mathrm{eff}T$), an abrupt change in $\alpha_\mathrm{eff}$ will noticeably affect $f_0$. Interestingly, two discontinuities are observed in Fig.~\ref{Fig:st-tempSweep}a: a deviation of $f_0$ below 165 K and a kink in $f_0$ at around 30 K (Fig.~\ref{Fig:st-tempSweep}a, inset). To identify the cause of these discontinuities, it is of interest to compare the temperatures at which they occur to the phase transitions in the bulk material.

Below 105 K, the cubic structure of bulk STO is known to break up into locally ordered tetragonal domains joined by ferroelastic domain walls~\cite{lytle1964x,unoki1967electron}. Another phase transition occurs at around 30 K, below which the Sr-ions disorder along [111] directions, rendering the structure locally triclinic~\cite{zalar05,scott12,salje13}. Both transitions are accompanied by changes in mechanical properties~\cite{ledbetter90,scott97,balashova95,chen99sto,kityk00}, as well as in the thermal expansion coefficient of STO~\cite{tsunekawa84} and can therefore be related to the signatures at 165 K and 30 K in Fig.~\ref{Fig:st-tempSweep}a. The difference between the observed elevated transition temperature of 165 K compared to the bulk value of 105 K has been attributed to surface effects~\cite{mishina00} and thermally induced strain~\cite{haeni04}. Additional evidence for an elevated cubic-to-tetragonal transition temperature is found from second harmonic generation measurements that we performed on the layers, which show a feature around 157 K (see Supplementary Figure 4). The fact that the resonance frequency increases with decreasing temperature, despite the decrease of the Young's modulus of bulk STO below the transition temperature~\cite{scott97,kityk00} indicates that the mechanical behaviour of the resonator is dominated by tension, rather than bending rigidity~\cite{castellanos13}.

While the phase transitions only lead to relatively small shifts in the resonance frequency (Fig.~\ref{Fig:st-tempSweep}a), they greatly impact the measured mechanical dissipation, as shown in Fig.~\ref{Fig:st-tempSweep}b. To characterise dissipation we use the quality factor of the resonator, which is extracted from the frequency domain measurements as $Q = f_0/\Delta f$ ($\Delta f$ is the full width at half maximum of the resonance peak). Whereas $f_0$ is influenced both by the pre-tension of the membrane and the Young's modulus of the STO, the quality factor is also dependent on the intrinsic losses in the material ($Q\propto E_1/E_2$ for a complex Young's modulus $E = E_1+iE_2$~\cite{unterreithmeier10}). In Fig.~\ref{Fig:st-tempSweep}b we show the evolution of the $Q$ factor for 7 alternating upward and downward temperature sweeps. In the regions below 30 K and above 165 K the data show a continuous temperature dependence. At intermediate temperatures, two distinct branches are observed: a low-dissipation branch, traced out by the red data points (belonging to the first two temperature sweeps, including the measurement from Fig.~\ref{Fig:st-tempSweep}a), and a high-dissipation branch traced out by the blue data points. The light blue points correspond to measurements taken during cooling down and the red and dark blue points are measurements taken during warming up. The overlap between the measurements in both directions implies that there is no correlation between the dissipation state and the direction of the temperature sweep. Besides an upward shift of resonance frequency after the first temperature cycle, in all subsequent measurements the $f_0$-$T$ curves are reproducible with only small variations (1 MHz, see Supplementary Figure 5). This suggests that the effect responsible for the different $Q$ branches does not influence the mechanical stiffness of the nanodrum and is not correlated to or caused by a change in the resonance frequency. Subsequent measurements (shown in Supplementary Figure 5) indicate that dissipation states in between the red and the blue data points are also accessible, but values of $Q$ outside of the range defined by the measurements in Fig.~\ref{Fig:st-tempSweep}b were not measured. Similar trends in $f_0$ and $Q$ as a function of temperature were observed in two other samples (see Supplementary Figure 6).

\section{Discussion}

The trend of decreasing dissipation (increasing $Q$) at lower temperatures is often observed in 2D materials~\cite{chen09,morell16,will17} and in microelectromechanical systems in general~\cite{kim08} and is a subject of ongoing discussion. A proposed explanation for this effect is the increased in-plane tension which is known to lower dissipation in nanomechanical structures~\cite{verbridge06,unterreithmeier10,cartamil15,norte16}. However, the different dissipation states of the STO drums between 30 and 165 K are, to our knowledge, a phenomenon that has not been observed in other nanomechanical systems. Despite experimental efforts, we have not been able to systematically control the dissipation state which the system exhibits. Even under identical experimental conditions, different dissipation states are observed. Some key observations out of 14 temperature sweeps on the same sample are: i) the low dissipation (high $Q$) state could only be observed after cooling down below 30 K and then warming up; ii) a transition from the low- to the high-dissipation state was observed once at around 85 K, after which the system was locked in the high dissipation branch; iii) the only way to get back to the low dissipation branch was to warm up above 160 K and then cool it down to 5 K.

From these observations it appears that the low-dissipation state is less stable than the high-dissipation state. The appearance of multiple phase states is reminiscent of the behaviour of phase-change materials that can enter crystalline, poly-crystalline or amorphous phases triggered by slight changes in cooldown rate. These phases can have huge variations in electrical resistance as well as in mechanical and optical properties. A potential comparable microscopic candidate for such behaviour in \ce{SrTiO3} could be its ferroelastic domain structure, since the motion of the domain walls has previously been linked to mechanical dissipation~\cite{salje13}. Nevertheless, to confirm this hypothesis, a simultaneous measurement of the mechanical properties and the domain wall configuration is needed.

\begin{figure}[t!]
	\centering
	\includegraphics[width=7cm]{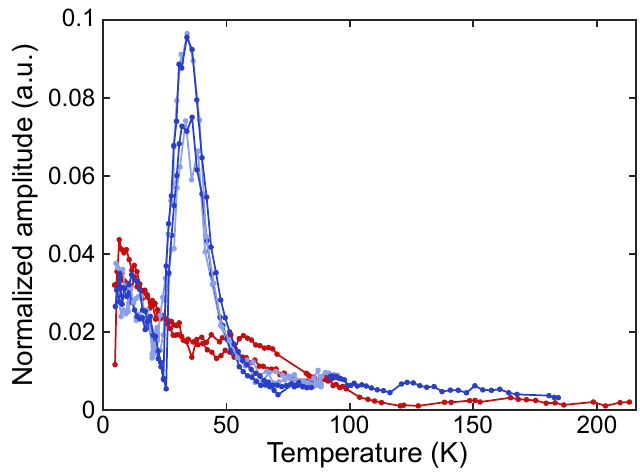}
	\caption{\label{Fig:st-amplitude} \textbf{Normalized motion amplitude at resonance.} The y-axis shows the resonance peak amplitude divided by the dc voltage on the photodiode to compensate for changes in the reflectivity of the sample. The blue and red points correspond to the data from Fig.~\ref{Fig:st-tempSweep}b.}
\end{figure}
	
Finally we discuss the measured amplitude at resonance, presented in Fig.~\ref{Fig:st-amplitude}. The amplitude at resonance $|A|_\mathrm{\omega=\omega_0}$ decreases with increasing temperature, which can be explained by the significant decrease in the quality factor (observed in Fig.~\ref{Fig:st-tempSweep}b) because in a harmonic oscillator $|A|_\mathrm{\omega=\omega_0} = FQ/(\omega_0^2m)$, where $F$ is the driving force, $\omega_0$ is the natural frequency and $m$ is the effective mass of the resonator. Interestingly, the measurement points that belong to the high-dissipation branch (blue points in Figures~\ref{Fig:st-tempSweep}b and~\ref{Fig:st-amplitude}) exhibit a large peak at around 34 K. This is counterintuitive, as one would expect that a lower quality factor should result in a lower amplitude. However, it might be that the driving force $F$ is temperature dependent and more than compensates for the decrease in the $Q$-factor. If so, this phenomenon cannot be caused by a change in the optical absorption of STO (at $\lambda~=~405$~nm), which is constant as a function of temperature in the measured range~\cite{kok15}.  Therefore, it is more likely that the increased driving efficiency is caused by a difference in the thermal properties between the low- and high-dissipation states of the material. In connection to this, it is worth mentioning that a peak in the heat capacitance of bulk \ce{SrTiO3} has been previously observed in a similar temperature window~\cite{duran08}.


In conclusion, we demonstrated the fabrication of ultrathin mechanical resonators made of epitaxially grown STO and SRO films. Using laser interferometry, we mechanically characterised the nanodrums and showed that they can be used as nanomechanical devices, much like drums made of van der Waals materials~\cite{bunch07,castellanos13,wang15,cartamil15,morell16,cartamil17BN}. We show that phase transitions affect the temperature-dependent dynamics of the resonators and that their mechanical dissipation can shed light on the microscopic loss mechanisms, which are often coupled to electronic and magnetic degrees of freedom.
This work connects and presents advances in two fields: (i) the field of complex oxides will benefit from a method for probing the mechanical properties of these strongly correlated electron materials in suspended form; (ii) the field of nanomechanics will now have access to a class of atomically-engineerable materials and heterostructures with exotic properties that can be used as functional elements in NEMS. Such nanomechanical resonators can be used in self-transducing mechanical devices, suspended Bragg reflectors, bimorphic actuators and novel thermomechanical and piezoelectric sensors. Furthermore, by decoupling the high-temperature growth of the materials from the device fabrication flow, these complex oxide NEMS can be easily integrated into fully functional CMOS devices, that cannot tolerate temperatures above 400$^\circ$C.

	
\section{Methods}
\label{st-methods}

\subsection{Pulsed laser deposition of epitaxial films}

SAO, STO, and SRO films were grown by pulsed laser deposition on \ce{TiO2}-terminated STO(001) substrates. The pulses were supplied by a \ce{KrF} excimer laser and the substrate was mounted using two clamps and heated by an infrared laser. SAO and STO were deposited using a laser fluence of $1.2\;\mathrm{J}/\mathrm{cm}^2$, a substrate temperature of $850^\circ\mathrm{C}$ and an oxygen pressure of $10^{-6}\;\mathrm{mbar}$. SRO was deposited at $600^\circ\mathrm{C}$, with a fluence of $1.1\;\mathrm{J}/\mathrm{cm}^2$ and an oxygen pressure of $0.1\;\mathrm{mbar}$. The growth occurred in layer-by-layer mode for SAO and STO, while SRO was grown in step-flow mode. After the deposition, the heterostructures were annealed for one hour at $600^\circ\mathrm{C}$ in 300 mbar \ce{O2}, and cooled down in the same atmosphere.
	
\subsection{Release \& transfer}

The thin films were released by adhering a PDMS layer to the film surface and immersing the stack in water. Dissolution of a $50\;\mathrm{nm}$ SAO layer was found to take approximately 60 minutes, without stirring or heating the water. After releasing the substrate, the PDMS layer with the thin film was dried using dry \ce{N2}. The STO and SRO films were transferred onto pre-patterned \ce{Si}/$285\;\mathrm{nm}$ \ce{SiO2} substrates using an all-dry deterministic transfer technique~\cite{castellanos14}. The crystallinity of the thin films before and after their release was investigated by X-ray diffraction (see Fig.~\ref{Fig:st-Char}b).

\subsection{Mechanical characterization}

The mechanical characterisation of the resonators (Figures 2-4) was performed with an active position feedback and variable frequency range, to ensure that the laser spot is always centered and focused on the drum. The resonance peaks are recorded with high accuracy (5000 points per measurement) to rule out any measurement artifacts in the interpretation of the data. In order to eliminate potential artifacts stemming from variations in the adhesion between the membranes and the substrate, the samples are thermally cycled prior to the measurement.

\subsection{Second harmonic generation}

The second harmonic generation (SHG) measurement was performed in a reflection geometry to further confirm the presence of the structural transition seen in the mechanical experiments. The sample was excited by a 60 fs laser pulse at a central wavelength of 800 nm from a regenerative Ti:Sapphire amplified laser system operating at a 1 kHz repetition rate. The fluence of the laser radiation used in the experiment was in the order of 10 mJ/cm$^2$.  The nonlinear response at the central wavelength of 400 nm was detected using a photomultiplier tube. 

\subsection*{Data availability}
\noindent The data that support the findings of this study are available from the corresponding authors upon reasonable request.

\section*{Acknowledgements}
The authors thank Pavlo Zubko and Gustau Catalan for the fruitful discussions and extensive feedback. This work was supported by the Netherlands Organisation for Scientific Research (NWO/OCW), as part of the Frontiers of Nanoscience (NanoFront) program, by the Dutch Foundation for Fundamental Research on Matter (FOM), by the European Union Seventh Framework Programme under grant agreement $\mathrm{n{^\circ}~604391}$ Graphene Flagship, and by the European Research Council under the European Union’s H2020 programme/ERC Grant Agreement No.~[677458].

\section{Author contributions}
D.J.G., A.M.V.R.L.M., and A.D.~deposited and characterised the epitaxial heterostructures and prepared the suspended films. D.D.~and D.J.G.~performed the measurements and analysed the data. D.D., D.J.G., A.M.V.R.L.M., H.S.J.~v.~d.~Z., A.D.C.~and P.G.S.~interpreted the data. D.A.~performed the second harmonic generation measurements. Y.~H. and E.~v.~H. synthetized the SAO target for pulsed laser deposition. P.G.S.~and A.D.C.~supervised the overall project. D.D., D.J.G., A.D.C~and P.G.S.~wrote the manuscript with input from all authors.

\section{Additional information}
\noindent Supplementary Information is available in the online version of the paper. Reprints and permissions information is available online at www.nature.com/reprints. Correspondence and requests for materials should be addressed to D.~D.~or D.~J.~G..

\section{Competing financial interests}
\noindent The authors declare no competing financial interests.

\pagebreak
\onecolumngrid
\setcounter{figure}{0}
\renewcommand{\figurename}{FIG. S\!\!}

\newpage
\section*{Supporting Information: Ultrathin complex oxide nanomechanical resonators}
\subsection{Growth and structural characterisation}
	
\begin{figure}[H]
		\centering
		\includegraphics[width=10cm]{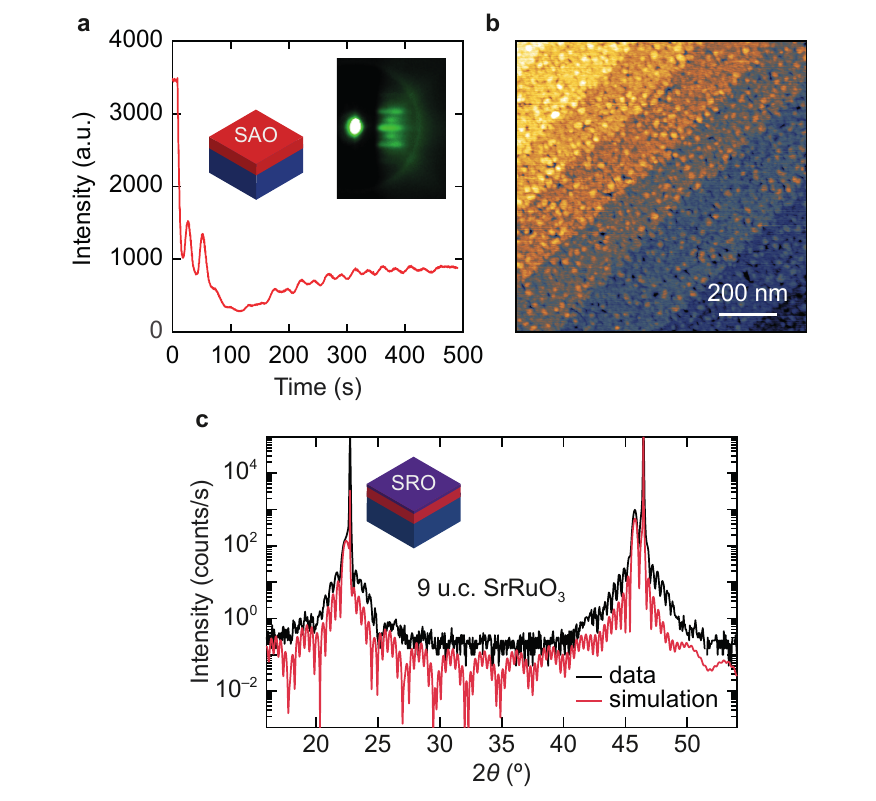}
		\caption{\label{Fig:char-supp}\textbf{Growth and characterisation.} \textbf{a}, RHEED oscillations during the growth of 19 u.c.~SAO on STO. Inset: RHEED diffraction pattern of the SAO surface. \textbf{b}, AFM topographic image of the SAO surface. \textbf{c}, XRD measurement of an SRO/SAO/STO heterostructure.}
	\end{figure}

	\noindent Figure \ref{Fig:char-supp}a shows RHEED oscillations during the growth of a 19 u.c.~SAO film on a \ce{TiO2}-terminated STO(001) substrate. An AFM measurement of the SAO surface is taken directly after the sample is removed from the vacuum chamber (see Fig.~\ref{Fig:char-supp}b). Clear steps and terraces are visible, indicating that the film grows in layer-by-layer mode. An XRD measurement of an SRO/SAO/STO heterostructure is shown in Fig.~\ref{Fig:char-supp}c. The simulation of the diffracted intensity (red line, calculated using InteractiveXRDFit~\cite{lichtensteiger2018interactivexrdfit}) can capture both the peak position and Laue oscillations of the SAO layer and the ultrathin (9 u.c.) SRO film.\\
	
	\noindent A recording of the release of an STO film on PDMS is included as a Supplementary Video.
	
	\newpage
	\subsection{Peak-force atomic force microscopy of suspended STO}
	
\begin{figure}[H]
		\centering
		\includegraphics[width=8.3cm]{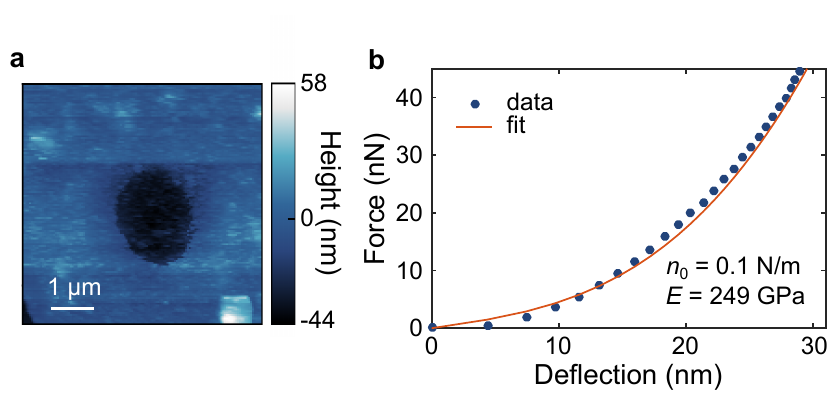}
		\caption{\label{Fig:st-afm} \textbf{Peak-force AFM of a suspended 4 $\mu$m STO drum.} \textbf{a,} Height profile. \textbf{b} Force-deflection curve taken in the centre of the drum. The red line is the fitted model, from which the Young's modulus and pre-tension are extracted.}
	\end{figure}

	\noindent In Fig.~\ref{Fig:st-afm} we use peak-force atomic force microscopy (AFM) to extract the pre-tension and the Young's modulus of the STO drum shown in Fig.~2a. The resonance frequency in the cross-over regime can be approximated as~\cite{castellanos2013single}:
	\begin{equation}
	\label{eq:crossover}
	f_0 = \sqrt{f_\mathrm{0,membrane}^2 + f_\mathrm{0,plate}^2}
	\end{equation}
	
	\noindent where $f_\mathrm{0,membrane} = \frac{2.4}{\pi d}\sqrt{\frac{n_0}{\rho h}}$ and $f_\mathrm{0,plate} = \frac{10.21}{\pi}\sqrt{\frac{E}{3\rho(1-\nu^2)}}\frac{h}{d^2}$. $E$ is the Young's modulus, $n_0$ the initial pre-tension (in N/m), $h$ the thickness, and $d$ is the diameter. Using the numbers obtained from Fig.~\ref{Fig:st-afm}, we get $f_0 = 19.4$ MHz, similar to the measured value for the same drum (see Fig.~2a).
	
	\newpage
	
	\subsection{Statistics of STO nanodrums}
	\label{st-hist}
	
\begin{figure}[H]
		\centering
		\includegraphics[width=12.7cm]{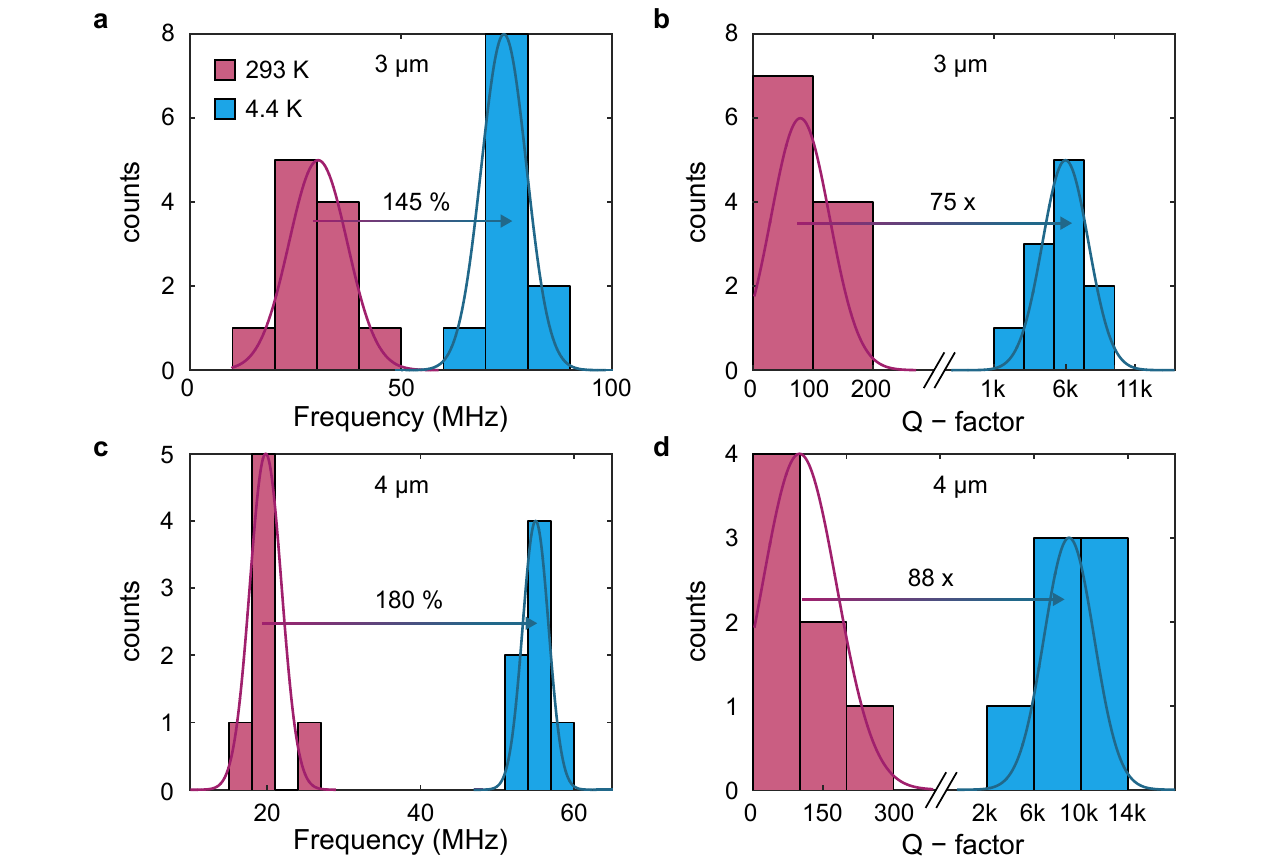}
		\caption{\label{Fig:histograms} \textbf{Statistical comparison of many STO drums between room- and low-temperature.} Histograms of the resonance frequencies of \textbf{a,} 3 $\mu$m and \textbf{c,} 4 $\mu$m STO drums. Histograms of the quality factor of \textbf{b,} 3 $\mu$m and \textbf{d,} 4 $\mu$m STO drums.}
	\end{figure}

	\noindent In Fig.~\ref{Fig:histograms} we show statistical data of eleven 3-$\mu$m and seven 4-$\mu$m drums measured at room temperature and at 4.4 K. In all the drums we see a drastic increase in the quality factor at low temperatures: an average of 75-fold increase in the 3-$\mu$m drums and 88-fold increase in the 4-$\mu$m drums. The resonance frequency is affected more strongly in the 4-$\mu$m drums, which is expected, because for larger drums the pre-tension has a stronger influence on the dynamic behaviour.

	\newpage
	\subsection{Second harmonic generation measurement of free-standing STO}
	\label{st-shg}
	
	Second harmonic generation (SHG) is a nonlinear optical technique based on the conversion of two photons of frequency $f$ to a single photon of frequency 2$f$. SHG has been shown to be an efficient method to probe microscopic transformations of the crystal symmetry near structural phase transitions. Both the polarization and the intensity of the SHG light can be observed when the crystal structure changes from between two symmetry groups at the phase transition point.
	
\begin{figure}[H]
		\centering
		\includegraphics[width=6.35cm]{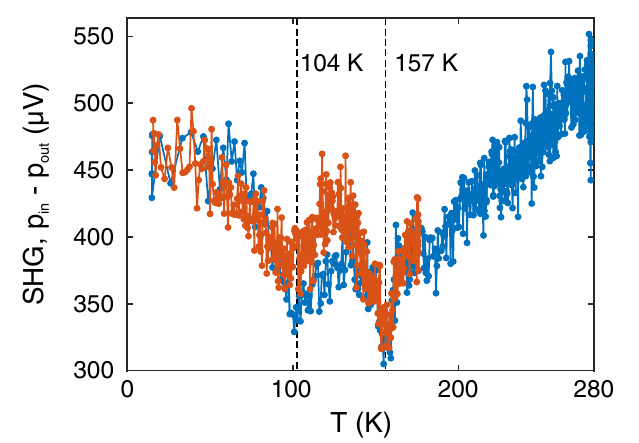}
		\caption{\label{Fig:shg} \textbf{SHG signal from the STO film.} SHG signal using a $p_\mathrm{in} - p_\mathrm{out}$ beam configuration taken while warming up (orange) and cooling down (blue).}
	\end{figure}

	\noindent An SHG measurement was performed on a 20 nm thick STO film transferred on a Si/\ce{SiO2} substrate. The measurement area is determined by the laser spot size in the setup, which is estimated to be around 100 $\mu$m. The polarization of the incoming light wave was controlled using a Glan-Taylor polarizer and a half-wave plate mounted in the rotation stage. The residual SHG signal from the optical components was filtered using a low-pass filter. The polarization of the reflected light was analyzed with a Glan-Taylor polarizer mounted in the rotational stage. The narrow band (10 nm) at the central filter was used to filter out light at the fundamental wavelength of 800 nm. The polarization of the incident light ($P_{in}$) is the same as the polarization of the recorded SHG signal ($P_{out}$). The measurements were taken during cooling down (blue curve in Fig.~\ref{Fig:shg}) and during warming up (orange curve in Fig.~\ref{Fig:shg}). The measurements show two features, one at 105 K, which corresponds to the transition temperature of bulk STO, and another one at 157 K. 
	
	\newpage
	
	\subsection{Full set of measurements on the STO resonator}
	\label{st-fulldata}
	
\begin{figure}[H]
		\centering
		\includegraphics[width=12.7cm]{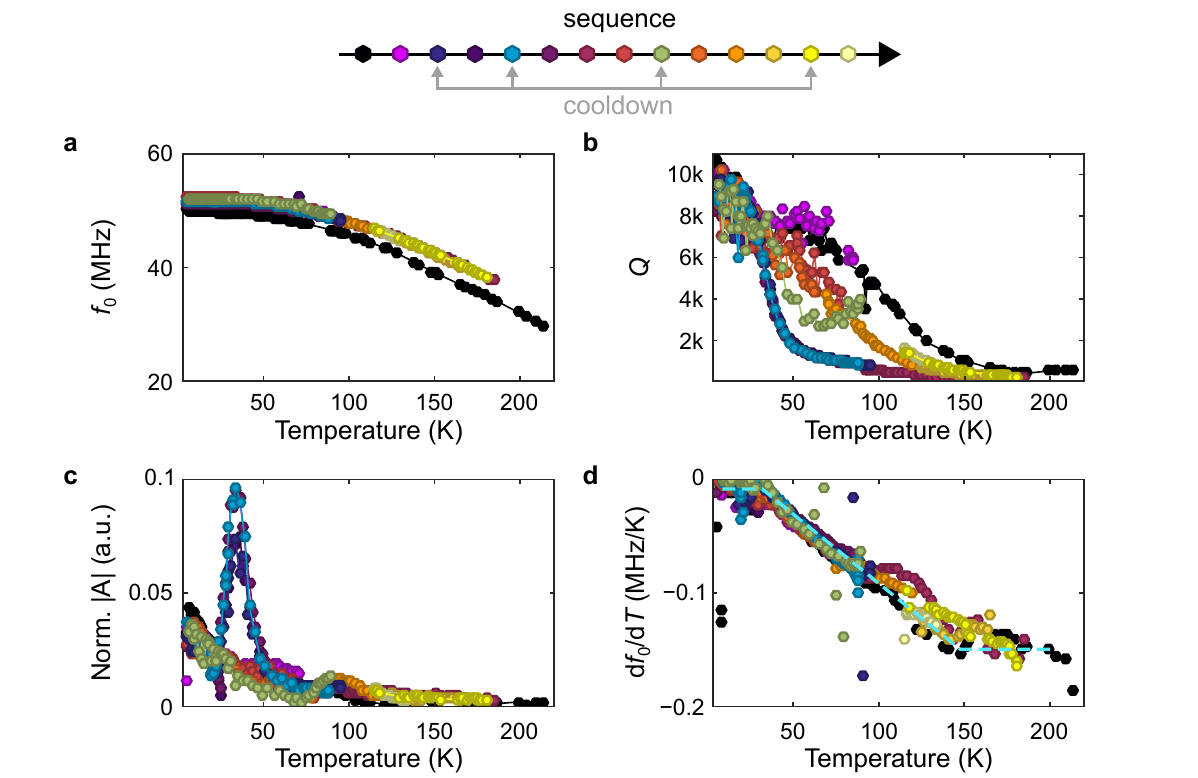}
		\caption{\label{Fig:full} \textbf{Temperature dependence for the STO drum from Fig.~3 in the main text for all 14 runs.} \textbf{a,} Resonance frequency, \textbf{b,} Q-factor, \textbf{c,} Normalised peak amplitude and \textbf{d,} d$f_0/$d$T$ vs. temperature. The measurement sequence is given above. The dashed blue line in \textbf{d} is a guide to the eye.}
	\end{figure}

	\noindent Figure~\ref{Fig:full} shows the full measurement set for the device shown in Fig.~2a, comprising 14 runs. Figure~\ref{Fig:full}d shows the derivative of the resonance frequency with respect to temperature, which is flat below 30 K and shows a change in slope above 150 K (illustrated by the dashed blue line in Fig.~\ref{Fig:full}d).
	
	\newpage
	
	\subsection{Temperature-dependent measurements on additional STO drums}
	\label{st-otherdevices}
	
\begin{figure}[H]
		\centering
		\includegraphics[width=12.7cm]{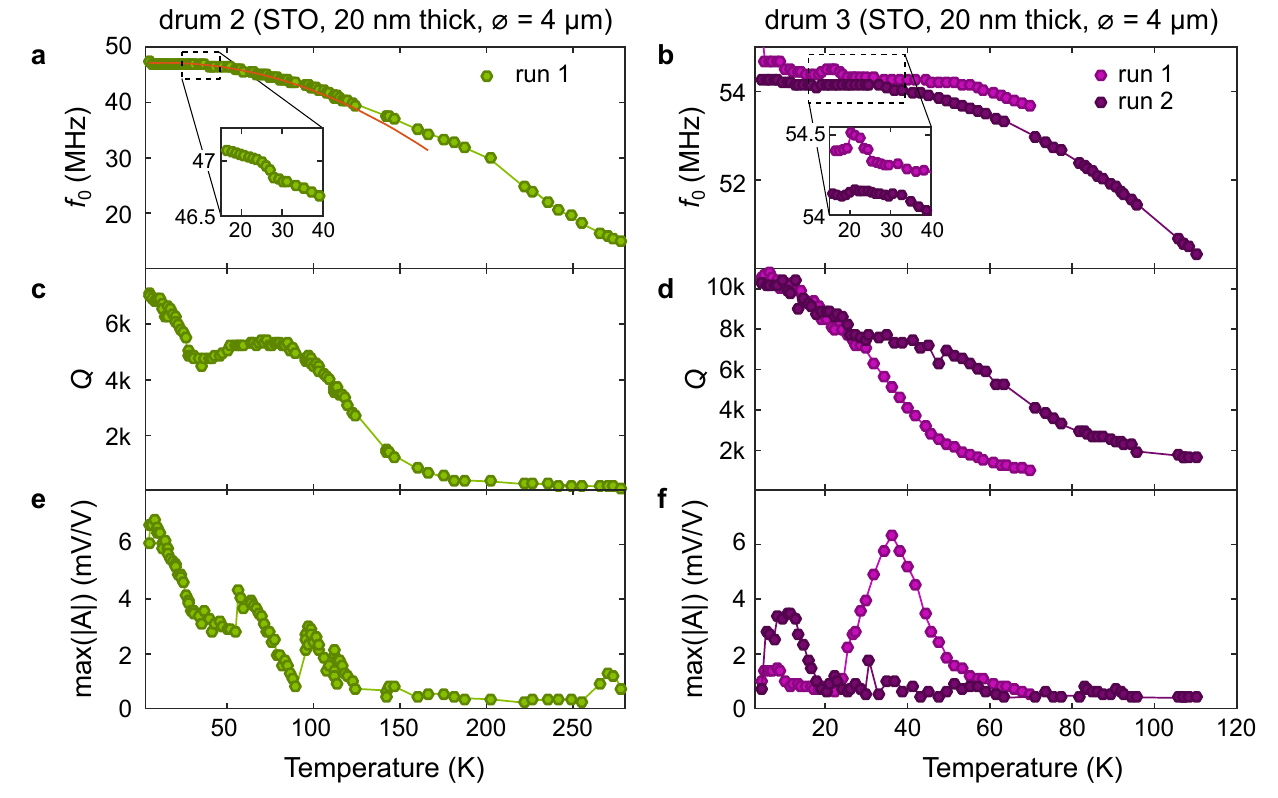}
		\caption{\label{Fig:otherdrums} \textbf{Temperature dependence for two other STO drums.} Resonance frequency vs. temperature for \textbf{a,} drum 2 and \textbf{b,} drum 3. The red curve in \textbf{a} is a guide to the eye. The insets show a zoomed-in region around 30 K. Q-factor vs. temperature for \textbf{c,} drum 2 and \textbf{d,} drum 3. Peak amplitude (not normalised) vs. temperature for \textbf{e,} drum 2 and \textbf{f,} drum 3. }
	\end{figure}

	\noindent Temperature-dependent measurements were performed on two additional STO drums (20 nm thick, 4 $\mu$m in diameter), shown in Fig.~\ref{Fig:otherdrums}. Drum 2 shows a sudden change of behaviour of the resonance frequency (Fig.~\ref{Fig:otherdrums}a), similar to the drum shown in Fig.~3, but at a lower temperature. Both samples show anomalies around 30 K, as shown in the insets of Fig.~\ref{Fig:otherdrums}a-b. The $Q$-factor of drum 3 also shows two distinct dissipation branches for the two temperature sweeps, plotted in different colours in Fig.~\ref{Fig:otherdrums}b,d,f. The low $Q$ branch is, as in the case for the main device, accompanied by an increase in the vibrational amplitude, as can be seen in Fig.~\ref{Fig:otherdrums}f. It is important to point out that the amplitude plotted in Fig.~\ref{Fig:otherdrums}e-f is not normalised with respect to the photodiode voltage, so any effects stemming from changes in the optical properties of the material for the wavelength of the measurement laser ($\lambda$ = 632.8 nm) are not compensated in these plots.
	
\end{document}